\documentclass[showpacs,floatfix,superscriptaddress,showpacs,twocolumn,amssymb,amsfonts,prb,aps]{revtex4}
\usepackage{longtable,graphicx,epsfig,dcolumn}

\begin{document}
\bibliographystyle{revtex}
\title
{Surface Layering in liquids: The role of surface tension.}

\author{Oleg~Shpyrko}
\affiliation{Department of Physics, Harvard University, Cambridge
Massachusetts 02138 (USA)}

\author{Masafumi~Fukuto}
\affiliation{Department of Physics, Harvard University, Cambridge
Massachusetts 02138 (USA)}

\author{Peter~Pershan}
\affiliation{Department of Physics, Harvard University, Cambridge
Massachusetts 02138 (USA)}

\author{Ben~Ocko}
\affiliation{Department of Physics, Brookhaven National Lab, Upton
New York 11973 (USA)}

\author{Ivan~Kuzmenko}
\affiliation{CMC-CAT, Argonne National Lab, Argonne, Illinois
60439 (USA) }

\author{Thomas~Gog}
\affiliation{CMC-CAT, Argonne National Lab, Argonne, Illinois
60439 (USA) }

\author{Moshe~Deutsch}
\affiliation{Department of Physics, Bar-Ilan University, Ramat-Gan
52900 (Israel)}

\date{Received 30 December 2003; revised manuscript received 25
March 2004}

\begin{abstract}

Recent measurements show that the free surfaces of liquid metals
and alloys are always layered, regardless of composition and
surface tension, a result supported by three decades of
simulations and theory. Recent theoretical work claims, however,
that at low enough temperatures the free surfaces of \textit{all}
liquids should become layered, unless preempted by bulk freezing.
Using x-ray reflectivity and diffuse scattering measurements we
show that there is no observable surface-induced layering in water
at T=298 K, thus highlighting a fundamental difference between
dielectric and metallic liquids. The implications of this result
for the question in the title are discussed.

\end{abstract}


\maketitle

\begin{figure}[tbp]
\epsfig{file=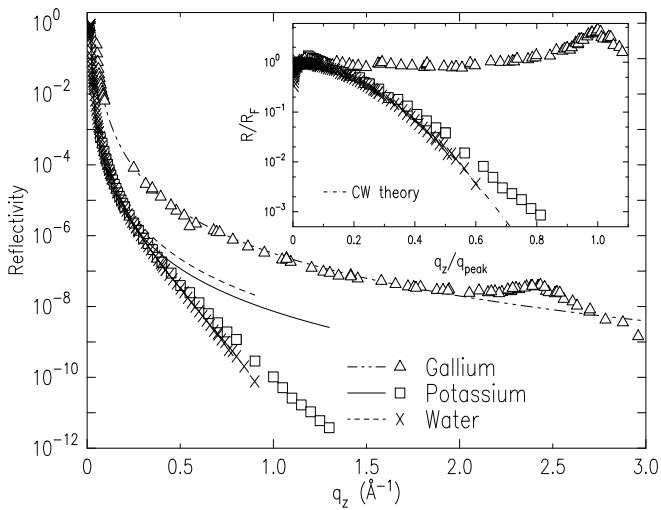, angle=90, width=1.0\columnwidth}
\caption{\label{fig1} X-ray reflectivity from free liquid surfaces
of the indicated samples. Points - measured [$R(q_z)$], lines -
calculated for an ideally flat and steplike interface
[$R_F(q_z)$]. The inset shows the ratio of the two. }
\end{figure}

The free surface of liquid metals and alloys were demonstrated
experimentally over the last few years to be layered, i.e. to
exhibit an atomic-scale oscillatory surface-normal density
profile.\cite{Magnussen95, Regan95,Tostmann00,Yang99} This is
manifested by the appearance of a Bragg-like peak in the x-ray
reflectivity (XR) curve, $R(q_z)$, as shown in Fig. \ref{fig1} for
Ga.\cite{Regan95} The wave-vector transfer position of the peak,
$q_{peak}$ is related to the layering period $d$ by
$q_{peak}=2\pi/d$. The layered interface is in a marked contrast
with the theoretical description of the liquid-vapor interface of
a simple liquid. This theory, prevailing for over a century,
depicts the density profile as a monotonic increase from the low
density of the vapor and the high density of the bulk
liquid.\cite{Buff65,Rowlinson82} This view was supported by XR
measurements on many non metallic liquids measured over the last
two decades, including water,\cite{Braslau85} alkanes,
\cite{Ocko94} and quantum liquids, \cite{Lurio92} which showed no
Bragg-like peaks. However, as discussed below, the measurements in
all of these studies were restricted to the small q range, i.e.
$q_z \ll \pi/$atomic size, and would not have detected surface
layering if it existed.

Early simulations on non-metallic liquids demonstrate that atomic
layering is ubiquitous near a hard flat surface
 and this has been observed for liquid Gallium.\cite{Huisman97}
On the other hand, for the liquid  vapor interface it is tempting
to think that the large surface tension $\gamma$ of liquid metals
such as Hg ($\gamma\approx500$ mN/m), Ga ($\gamma\approx750$
mN/m), and In ($\gamma\approx550$ mN/m) might be the explanation
for the SL observed at their surface. This assumption is partially
mitigated by the observation of SL at the free surface of liquid K
where $\gamma\approx100$ mN/m only. Here we report x-ray
scattering results showing that the free surface of water, which
has nearly the same surface tension as K, does not exhibit SL
features in the reflectivity profiles, thereby suggesting that
surface tension by itself does not explain SL. This conclusion
rests on the validity of the capillary wave theory (discussed
below).

Rice \textit{et al}. \cite{Rice74} first predicted SL in liquid
metals three decades ago. They argued that the layered interface
structure for liquid metals is a consequence of the strong
dependence of the effective ion-potential energy on the steeply
varying electron density across the liquid/vapor interface. At the
low density vapor phase the electrically neutral atoms interact
through a van der Waals interaction only. In the metallic, higher
density, liquid phase the electrons are delocalized, and the much
more complex interactions involve an interplay between a quantum
Fermi fluid of free electrons and a classical liquid of charged
ion cores. Rice concludes that this substantive change in the
effective-ion-potential stabilizes the short range surface
fluctuations with the result that the atoms near the surface form
a layered structure. \cite{Rice74} Calculations employing the glue
model of metallic cohesion support these conclusions.
\cite{Celestini97} By contrast, Soler \textit{et
al.}\cite{Soler01} claim that SL is solely due to the formation of
a dense layer at the surface. This layer is not restricted to
metallic liquids, but may form also in nonmetallic liquids due to
nonisotropic interactions, such as remnant covalent bonding in
liquid Si\cite{Soler01} and the highly directional interactions in
liquid crystals,\cite{Pershan82} or in non-uniform Lennard-Jones
fluids with unbalanced attractive forces.\cite{Katsov02}
Additionally, the SL has been found in colloidal systems,
\cite{Madsen01} which suggests that layering phenomenon may be
more fundamental than previously thought. More recently, based on
extensive simulations, Chac{\'o}n \textit{et al.}\cite{Chacon01}
argued that SL does not require the many-body, delocalize-electron
interactions of a liquid metal at all. Rather, according to them,
SL is a universal property of all liquids at low enough
temperatures, $T \lesssim T_c/a$, whenever not preempted by bulk
solidification. Here $T_c$ is the critical temperature of the
liquid, and $a \approx 4-5$.

The dichotomy between these two views could be resolved, in
principle, by XR measurements  on selected materials
\cite{Deutsch98} to see which exhibit, or not, a layering peak at
some $q_{peak}$. Unfortunately, practical considerations limit the
number of elemental liquids that can be studied to a relatively
few. One of the major problems is that the measurable $q_z$ range
is more often than not limited to values much less than $q_{peak}$
by the strong off-specular diffuse scattering  caused by thermal
capillary waves. The effect of the capillary waves is to induce a
surface roughness $\sigma \sim \sqrt{T/\gamma}$, where $\gamma$ is
the surface tension. The consequence of this, which is shown in
Fig.~\ref{fig1}, for three liquids at room temperature, is to
reduce the reflectivity $R(q_z)$ below that of the theoretical
Fresnel reflectivity from an interface with an idealized flat,
steplike surface-normal density profile. For low-$\gamma$ liquids
such as water ($\sim$ 70 mN/m) and K ($\sim$100 mN/m) the
reduction is significant. For Ga, where $\gamma \approx 750$ mN/m
the effect is almost negligible at room temperature. Even that
small reduction is almost completely offset at room temperature by
the SL effect that peaks at $q_z \approx 2.5~${\AA}$^{-1}$. At
higher temperatures, however, the effect is quite prominent.
\cite{Regan95}

Although the rapid falloff in $R/R_F$ of water prevents its
measurements out to $q_z\approx q_{peak}=2.0~\AA^{-1}$, our recent
studies of liquid K \cite{Shpyrko03} demonstrated that if surface
layering is present its signature can still be observed clearly at
$q_z<<q_{peak}$ even for low $\gamma\lesssim 100$\ mN/m liquids.
This is accomplished by carefully accounting for the effects of
capillary waves, based on diffuse x-ray scattering (DS)
measurements. We present here an x-ray study of the surface
structure of water over the most extended $q_z$-range published to
date, and including diffuse scattering \cite{Braslau85,
Schwartz90}. For the present measurements the intrinsic surface
structure factor of water can be extracted directly from the raw
$R(q_z)$ without resorting to any structural model for the
interface. Comparison between the water surface structure factor,
for which there is no evidence of SL, and that of K and Ga
suggests that surface tension is not the dominant cause of the SL
observed in liquid metals.

X-ray measurements were carried out on the CMC-CAT liquid surface
diffractometer, APS, Argonne National Laboratory, at a wavelength
of $\lambda=1.531~{\AA}$. The purified water sample was contained
in a Langmuir trough\cite{Schwartz92} mounted on the
diffractometer. The surface was periodically swept with a teflon
barrier, monitoring $\gamma$ with a film balance, to ensure a
clean surface.\cite{Gaines66}

\begin{figure}[tbp]
\epsfig{file=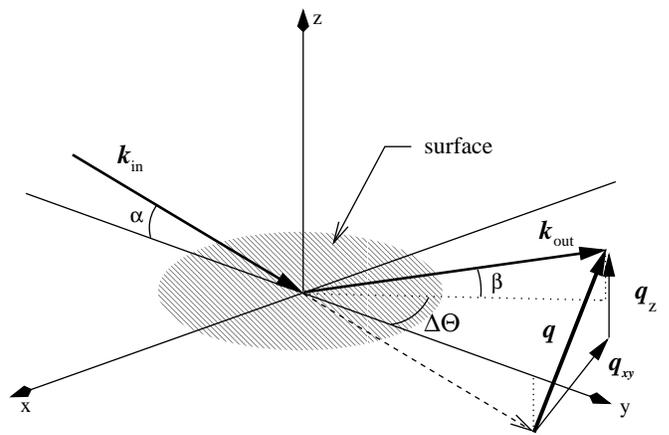, angle=0, width=1.0\columnwidth}
\caption{\label{fig2} Schematic description of the experimental
setup. }
\end{figure}
Both XR \cite{Deutsch98} and DS \cite{Pershan00} are well
documented techniques. Experimental geometry setup employed in the
experiments described in this work is shown in Fig.~\ref{fig2}.
For x-rays striking the surface at a grazing angle $\alpha$ and
detected at an output angle $\beta$ \textit{in the specular
plane}, the surface-normal ($z$) and in-plane surface-parallel
($y$) components of the wavevector transfer are $q_z = (2 \pi /
\lambda) \left( \sin \alpha + \sin \beta \right)$ and $q_{y} = (2
\pi / \lambda) \left( \cos \alpha - \cos \beta \right)$. In a XR
measurement both angles are varied, keeping $\alpha = \beta $, so
that $q_{y}=0$, and the reflected signal, measured vs $q_z$ and
divided by the incident intensity $I_0$, yields $R(q_z)$. In a DS
measurement, $\beta$ is varied for a fixed $\alpha$, and hence
both $q_z$ and $q_y$ vary.

Within the Born approximation,\cite{Deutsch98}
$R(q_z)/R_F(q_z)=|\Phi (q_z)|^2 W(\eta,q_z)$, where $\Phi (q_z) =
(\rho_{\infty})^{-1} \int [\textrm{d}\left< \rho(z)
\right>/\textrm{d}z] \exp(\imath q_z z) \textrm{d}z$\ is the
conventional structure factor of the liquid-vapor interface,
$\left< \rho(z) \right>$\ is the \textit{intrinsic} (i.e. in the
absence of capillary wave smearing) surface-parallel-averaged
electron density profile along the surface-normal $z$ direction,
and $W(\eta,q_z)$ accounts for the smearing of the intrinsic
density profile by capillary waves. The aim of our XR measurement
is to determine $\Phi (q_z)$ to observe a possible layering peak.
Since both $\Phi (q_z)$ and $W(\eta,q_z)$ depend on $q_z$,
extracting $\Phi (q_z)$ directly from the measured $R(q_z)$ is
possible as $|\Phi(q_z)|^2=R(q_z)/W(\eta,q_z)$ only if
$W(\eta,q_z)$ is known independently. For this we use the DS data.

\begin{figure}[tbp]
\epsfig{file=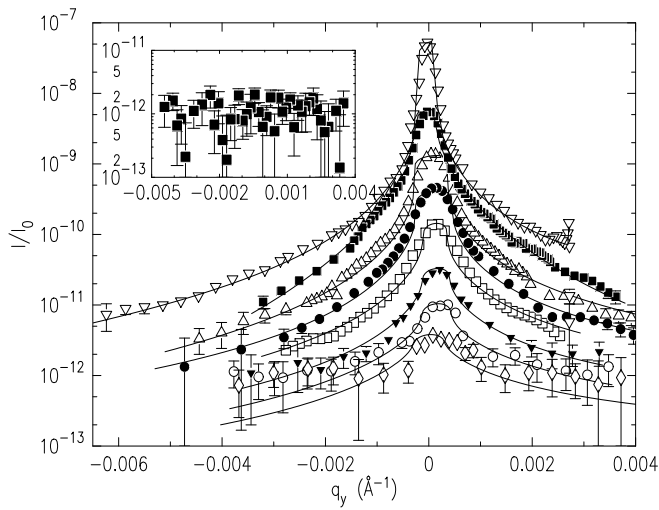, angle=90, width=1.0\columnwidth}
\caption{\label{fig3} Comparison of measured diffuse scattering
 with capillary wave theory predictions for the
angles of incidence $\alpha$ (top to bottom): 2.1$^{\circ}$,
2.8$^{\circ}$, 3.5$^{\circ}$, 4.2$^{\circ}$, 5.0$^{\circ}$,
5.7$^{\circ}$, 6.0$^{\circ}$, and 6.4$^{\circ}$. The last data set
corresponding to 7.1$^{\circ}$ shown in the inset no longer
exhibits a distinguishable specular peak. The $q_z$ values
corresponding to the specular condition $q_y=0$ \AA$^{-1}$ are
0.3, 0.4, 0.5, 0.7, 0.8, 0.85, 0.9, and 1.0 $\AA^{-1}$,
respectively. }
\end{figure}

The measured DS, shown in Fig.~\ref{fig3}, is given by theory as
$I_{DS}=\int[\textrm{d}\sigma(q_x-q'_x,
q_y-q'_y,q_z-q'_z)/\textrm{d}\Omega]\textrm{d}\omega(q'_x,q'_y,q'_z)$,
where $\textrm{d}\omega$ is the angular resolution of the
diffractometer and
\begin{equation}
\frac{d\sigma }{d\Omega } = \frac{A_{0}}{8\pi\sin \alpha } q_z ^2
R_F(q_z) \mid \Phi (q_z) \mid ^2 \frac{\eta}{q_{xy}^{2-\eta}}
\left( \frac{1}{q_{ max}} \right) ^\eta \: \label{eq:master}
\end{equation}
is the scattering cross-section. Here $q_c$ is the critical angle
for total external reflection of x rays, $q_{max} \approx \pi/\xi$
is the upper cutoff for capillary wave contributions, with $\xi$
of order of the atomic diameter, and $ \eta =(k_BT)/(2\pi
\gamma)q_z^2 $.

Two subtle complications are encountered in analyzing the DS data.
First, nonsurface DS contributions (scattering from the bulk,
sample chamber windows, etc.) can significantly distort the shape
of the DS scans. These background contributions are measured by
offsetting the detector by $\Delta\Theta = \pm 0.3^\circ$ from the
plane of incidence, and are already subtracted from the data shown
in Fig.~\ref{fig3}. Second, in the fixed-$\alpha$ DS scans,
carried out by scanning $\beta$, $q_z$, and thus $|\Phi(q_z)|^2$,
also vary. Fortunately, $\delta q_z \approx \delta q_y /\beta$ and
for the range of $q_y$ displayed in Fig.~\ref{fig3} and typical
$q_z$ values, the changes are small enough to be neglected, e.g. a
maximal $\delta q_z \approx 0.06~{\AA}^{-1}$\  at the largest
$\beta$\ for $\alpha=3.5^{\circ}$. Thus, it is a good
approximation for each of the DS scans to treat $\Phi(q_z)$ as a
fixed function of $\alpha$. $\mid\Phi(q_z )\mid^2$ is then
obtained by dividing the measured DS and XR curves by
$W(\eta,q_z)=\int[A_{0}/(8\pi\sin \alpha)] q_z ^2 R_F(q_z)
(\eta/q_{y}^{2-\eta})q^{-\eta}_{max} \textrm{d}\omega$.
\cite{Shpyrko03}

\begin{figure}[tbp]
\epsfig{file=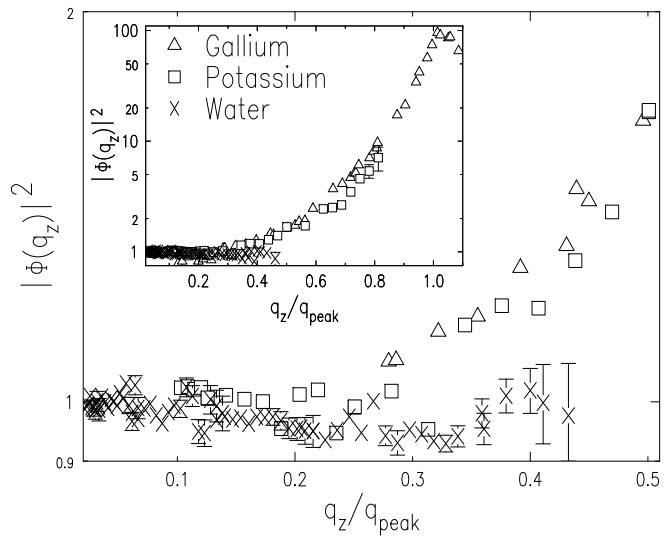, angle=90, width=1.0\columnwidth}
\caption{\label{fig4} Comparison of the structure factor squared
$\mid \Phi(q_z) \mid ^2$ for water (crosses), liquid potassium
(squares), and liquid gallium (triangles). The wavevector $q_z$ is
normalized to the expected position of the layering peak
$q_{peak}$ of each sample. The inset shows the data on an extended
scale. For discussion see text.}
\end{figure}

The theoretical curves calculated using Eq.~\ref{eq:master} with
$T=298~K$ and $\gamma=72~$mN/m are shown as lines in Fig.
\ref{fig3}. As $q_z$ increases, so do both $\eta \propto q_z^2$
and the intensity of the off-specular power-law wings relative to
that of the specular peak at $q_y=0$. The curve at $q_z=1$
\AA$^{-1}$ demonstrates the capillary-wave-imposed limit where the
specular signal at $q_y=0$, which contains the surface structure
information, becomes indistinguishable from the DS signal at $\pm
q_y > 0$. In principle, this limit arises from the fact that for
$\eta \geq 2$ the singularity at $q_{y}=0$ in $d\sigma/d\Omega$
vanishes and there is no longer any criterion by which the surface
scattering can be differentiated from other sources of diffuse
scattering. In practice, the fact that the projection of the
resolution function on the horizontal x-y plane is very much wider
transverse to the plane of incidence than within the plane of
incidence reduces this limit to a value closer to $\eta \approx
1$. \cite{Shpyrko03} Figure~\ref{fig3} exhibits excellent
agreement between the theoretical DS curves calculated from Eq.
\ref{eq:master} with the measured DS over several decades in
intensity and one decade in $\eta$, \textit{without any adjustable
parameters}. This confirms the applicability of the capillary wave
theory for the surface of water over the $q_z$ range studied here,
$0 \leq q_z \leq 0.9$\ \AA$^{-1}$. Measurements of diffuse
scattering for small $q_z$ (i.e. small $\eta$) to values of the
surface parallel component of the wave vector transfer $q_x$ of
the order of $\pi/$atomic size have been done by Daillant
\textit{et. al} \cite{Fradin00, Mora03} by moving the detector out
of the plane of incidence for grazing incident angles. At such
large wave vectors the observed scattering must be interpreted as
the superposition of scattering due to surface capillary waves and
bulk diffuse scattering from the liquid below the surface. The
separation of these two contributions is rather subtle and was
only accomplished through nontrivial calculations of the
noncapillary terms. Eventually Daillant \textit{et. al} concluded
that as the value of $q_x$ approaches the atomic scale (i.e.,
$\pi/$atomic size) the surface tension varies with
$q_x$.\cite{Mecke99} In principle, this dispersion should affect
the sum rule that is the origin of the $1/q_{xy}^\eta$ in the
differential cross section in Eq.(\ref{eq:master}). On the other
hand, considering that the dispersion is a relatively small effect
that will only change the argument of a logarithmic term, its
effect on the present analysis can be neglected.

$|\Phi(q_z)|^2$ is then obtained directly from the measured XR
curve, as $R(q_z)/W(\eta,q_z)$ as discussed above. It is shown in
Fig.~\ref{fig4} (circles) along with previously measured results
for K (squares) and Ga (triangles).

The rise of the Ga $|\Phi(q_z)|^2$ to $\sim$100 at $q_{peak}$ due
to layering can be clearly seen on the inset of Fig.~\ref{fig4}.
For K, the capillary-wave-imposed limit only allows obtaining
$|\Phi(q_z)|^2$ for $q_z \lesssim 0.8 q_{peak}$. Nevertheless, the
value of $|\Phi(q_z)|$ for K starts to deviate from unity for
values of $q_z/q_{peak} \approx 0.3$. Furthermore, over the range
for which it can be measured it is basically identical to the
structure factor of Ga. This is a clear indication that the
surface of liquid K has essentially the same SL as that of Ga,
which is also nearly identical to that of the other liquid metals
that have been studied to date, e.g., In\cite{Tostmann99} and
Sn\cite{Shpyrko03b}.  For water, however, no deviation of
$|\Phi(q_z)|^2$ from unity is observed even at the highest
measurable $q_z/q_{peak}\approx0.5$. This suggests that
surface-induced layering does not occur at the surface of water.
The different behavior, in spite of the similar $\gamma$ of water
and K, leads to the conclusion that the surface layering in K, and
by implication in other liquid metals, is not merely a consequence
of its surface tension.

The absence of layering in water, and its presence in potassium,
seems at first sight to corroborate the claim of Rice \textit{et
al.} \cite{Rice74} that layering is a property arising from the
metallic interaction of the liquid. On the other hand, Chac{\'o}n
\textit{et al.}\cite{Chacon01}, who maintain that surface-induce
layering is a general property of all liquids, regardless of their
interactions, predict that layering should occur only at
temperatures $T/T_c \lesssim 0.2$, where $T_c$ is the critical
temperature of the liquid. Although supercooling is often
possible, the practical limit for most reflectivity measurements
is the melting temperature $T_m$. Thus the smallest $T/T_c$ for
any liquid is on the order of $T_m/T_c$. For liquid metals
$T_m/T_c\approx 0.15$ (K), 0.13 (Hg), 0.07 (In), 0.066 (Sn), and
0.043 (Ga). Since these values are $ < 0.2$, by Chac{\'o}n's
criteria surface layering is expected, and indeed demonstrated
experimentally to occur, in all of
them.\cite{Magnussen95,Regan95,Tostmann99, Shpyrko03, Shpyrko03b}
By contrast, for water, where $T_m/T_c=0.42 \gg 0.2$, Chac{\'o}n's
criteria predict that the appearance of surface layering is
preempted by bulk freezing, and thus no SL should occur at room
temperature, as indeed found here.

In summary, we have shown that the surface of water does not
exhibit SL even though its surface tension is not significantly
different from that of liquid K for which SL was observed. Thus
the atomic layering of liquid K cannot be the result of surface
tension alone. Although this seems to support Rice's argument that
the metallic phase is essential for SL, we note that SL is
exhibited by both liquid crystals\cite{Pershan82} and other
intermediate-size organic molecules.\cite{Dutta99} Consequently,
there can be other criteria for SL, in addition to those proposed
by Rice. One possibility, proposed by Chac{\'o}n \textit{et. al},
is that SL should be ubiquitous for all liquids that can be cooled
to temperatures of the order of $0.2T_c$. Unfortunately, this is
difficult to explore experimentally since suitable liquids are
rather scarce. For example, liquid noble gases, the archetypical
van der Waals liquids, have all $T_m/T_c > 0.55$. For liquid
helium the ratio is lower but in view of both the presence of the
superfluid transition and the low scattering cross section this is
a difficult system to study. X-ray reflectivity measurements at
1~K, i.e., $T_m/T_c \sim 0.2$, did not exhibit evidence for
SL\cite{Lurio92}. Similarly, the polar liquids oxygen and fluorine
have $T_m/T_c$ of only 0.38 and 0.37. There are, however, some low
melting organic van der Waals liquids, e.g. propane and 1-butene,
that have $0.2 < T_m/T_c < 0.25$, and may be suitable for
addressing this issue. It would be also very interesting to
measure the surface structure of materials such as Se
($T_m/T_c=0.28$), Te (0.3), Sc (0.28), and Cd (0.22), for which
$T_m/T_c >0.2$. If it were possible to probe surface layering in
these high $T_m/T_c$ materials such experiments could test
Chac{\'o}n's argument. Unfortunately, these materials have
relatively high vapor pressure and that precludes the use of UHV
methods for insuring atomically clean surfaces. The surfaces of
some alkalis and mercury can be kept clean without UHV methods.
For most of the experimentally accessible metals this is not true.

This work has been supported by the U.S. Department of Energy
Grants Nos. DE-FG02-88-ER45379 and DE-AC0298CH10886, the National
Science Foundation Grant No. DMR-0124936 and the U.S.--Israel
Binational Science Foundation, Jerusalem. Work at the CMC
Beamlines is supported in part by the Office of Basic Energy
Sciences of the U.S. Dept. of Energy and by the National Science
Foundation Division of Materials Research. Use of the Advanced
Photon Source is supported by the Office of Basic Energy Sciences
of the U.S. Department of Energy under Contract No.
W-31-109-Eng-38.


\begin{thebibliography}{10}

\bibitem{Magnussen95} O.~M.~Magnussen, B.~M.~Ocko, M.~J.~Regan, K.~Penanen,
P.~S.~Pershan, and
  M.~Deutsch, Phys. Rev. Lett. \textbf{74}, 4444 (1995).



\bibitem{Regan95}  M.~J.~Regan, E.~H.~Kawamoto, S.~Lee, P.~S.~Pershan,
N.~Maskil, M.~Deutsch, O.~M.~  Magnussen, B.~M. Ocko, and L.~E.
Berman, Phys. Rev. Lett. \textbf{75}, 2498 (1995).

\bibitem{Tostmann00} H.~Tostmann, E.~DiMasi, P.~S. Pershan, B.~M. Ocko, O.~G.
Shpyrko, and M.~Deutsch, Physical Review B 61, 7284 (2000) ;
E.~DiMasi, H.~Tostmann, O.~G. Shpyrko, P.~Huber, B.~M. Ocko, P.~S.
Pershan, M.~Deutsch, and L.~E. Berman, Phys. Rev. Lett.
\textbf{86}, 1538 (2001).

\bibitem{Yang99} B.~Yang, D.~Gidalevitz, D.~X. Li, Z.~Q. Huang, and S.~A.
Rice, Proc. Nat. Acad. of Sc. USA \textbf{96}, 13009 (1999) ;
B.~Yang, D.~X. Li, Z.~Q. Huang, and S.~A. Rice., Phys. Rev. B
\textbf{62}, 13111 (2000).

\bibitem{Buff65} F.~P. Buff, R.~A. Lovett, and F.~Stilling, Phys. Rev. Lett.
\textbf{15}, 621 (1965) ; F.~H. Stillinger and J.~D. Weeks,  J.
Phys. Chem. \textbf{99}, 2807 (1995).

\bibitem{Rowlinson82} J.~S. Rowlinson and B.~Widom,
"Molecular Theory of Capilarity" (Oxford University Press, Oxford,
1982).

\bibitem{Braslau85} A.~Braslau, M.~Deutsch, P.~S. Pershan, A.~H. Weiss,
J.~Als-Nielsen, and J.~Bohr, Phys. Rev. Lett. \textbf{54}, 114
(1985).

\bibitem{Ocko94} B.~M. Ocko, X.~Z. Wu, E.~B. Sirota, S.~K. Sinha, and
M.~Deutsch, Phys. Rev. Lett. \textbf{72}, 242 (1994).



\bibitem{Lurio92} L.~B. Lurio, T.~A. Rabedeau, P.~S. Pershan, I.~F. Silvera,
M.~Deutsch, S.~D. Kosowsky, and B.~M. Ocko, Phys. Rev. Lett.
\textbf{68}, 2628 (1992) ; Phys. Rev. B \textbf{48}, 9644  (1993).

\bibitem{Huisman97} W.~J. Huisman, J.~F. Peters,M.~J. Zwanenburg et al., Nature
        \textbf{390}, 107, 379 (1997).

\bibitem{Rice74} S.~A.~Rice, D.~Guidotti, and H.~L.~Lemberg,
W.~C.~Murphy and A. N. Bloch, in "Advances in Chemical Physics
XXVII" edited by I. R. Prigogine, and S.~A. Rice (Wiley,
Chichester, 1974) p. 543 ; M. P. D'Evelyn and S. A. Rice, J. Chem.
Phys. \textbf{78}, 5225 (1983) ; S.~A. Rice, Mol. Simul.
\textbf{29}, 593 (2003).


\bibitem{Celestini97} F. Celestini, F. Ercolesi and E. Tosatti, Phys. rev.
Lett. \textbf{78}, 3153 (1997) ; S. Iarlori, P. Carnevali, F.
Ercolesi and E. Tosatti, Surf. Sci \textbf{211/212}, 55 (1989).

\bibitem{Soler01} G. Fabricius, E. Artacho, D. Sanchez-Portal, P. Ordejon,
D.~A. Drabold and J.~M. Soler, Phys. Rev. B (Rapid) \textbf{60},
R16283 (1999) ;  J.~M. Soler, G. Fabricius, and E. Artacho, Surf.
Sci. \textbf{482-485}, 1314 (2001).

\bibitem{Pershan82} P.~S. Pershan, Physics Today \textbf{35}(5), 34 (1982).

\bibitem{Katsov02} K.~Katsov and J.~D. Weeks, J. Phys. Chem. B \textbf{106},
8429 (2002).

\bibitem{Madsen01} A.~Madsen,O.~Konovalov, A.~Robert and
G.~Grubel, Phys. Rev. E \textbf{64}, 061406 (2001)

\bibitem{Chacon01} E.~Chacon, M.~Reinaldo-Falagan, E.~Velasco, and
P.~Tarazona, Phys. Rev. Lett. \textbf{87}, 166101 (2001) ;
E.~Velasco, P.~Tarazona, M.~Reinaldo-Falagan, and E.~Chacon, J.
Chem. Phys. \textbf{117}, 10777 (2002).

\bibitem{Deutsch98} M.~Deutsch and B.~M. Ocko, in "Encyclopedia of Applied
Physics", edited by G. L. Trigg (VCH, New-York, 1998), Vol. 23, p.
479 ; J. Als-Nielsen and D. McMorrow, "Elements of Modern X-Ray
Physics" (Wiley, New York, 2001).

\bibitem{Shpyrko03} O.~G. Shpyrko, P. Huber, A. Grigoriev, P.~S. Pershan, B.M.
Ocko, H. Tostmann,  and  M. Deutsch, Phys. Rev. B \textbf{67},
115405 (2003).



\bibitem{Schwartz90} D.~K. Schwartz, M.`L. Schlossman, E.`H. Kawamoto, G.~J.
Kellogg, P.~S. Pershan, and B.~M. Ocko, Phys. Rev. A \textbf{41},
5687 (1990).

\bibitem{Schwartz92} D. K. Schwartz, M. L. Schlossman, and P. S. Pershan,
J. Chem. Phys. \textbf{ 96}, 2356 (1992).

\bibitem{Gaines66} G. L. Gaines, "Insoluble Monolayers at Liquid-Gas
Interfaces" (Wiley, New York, 1966).

\bibitem{Pershan00} P. S. Pershan, Coll. Surf. A \textbf{171}, 149 (2000).

\bibitem{Fradin00} C. Fradin, A. Braslau, D. Luzet, D. Smilgies, M. Alba, N. Boudet, K.
Mecke, J. Daillant, Nature \textbf{403}, 871 (2000)

\bibitem{Mora03} S. Mora, J. Daillant, K. Mecke, D. Luzet, A. Braslau, M. Alba, B. Struth,
Phys. Rev. Lett. \textbf{90}, 216101 (2003).

\bibitem{Mecke99} K.~R.~Mecke and S.~Dietrich, Phys. Rev. E 59, 6766 (1999)

\bibitem{Tostmann99} H.~Tostmann, E.~DiMasi, P.~S. Pershan, B.~M. Ocko,
O.~G. Shpyrko, and
  M.~Deutsch, Phys. Rev. B \textbf{59}, 783 (1999).

\bibitem{Shpyrko03b} O.~G. Shpyrko, A.~Grigoriev, P.~S. Pershan, C.~Steimer,
B.M. Ocko, M.~Deutsch,   B.~Lin, M.~Meron, T.~Graber, and
J.~Gerbhardt, ( to be published).

\bibitem{Dutta99}
C.~J. Yu, A.~G. Richter, A. Datta, M.~K. Durbin and P. Dutta,
Phys. Rev. Lett. \textbf{82}, 2326 (1999).

\end{thebibliography}

\end{document}